# An Empirical Study on Refactoring Activity

Mohammad Iftekharul Hoque, Vijay Nag Ranga, Anurag Reddy Pedditi, Rachitha Srinath,
Md Ali Ahsan Rana, Md Eftakhairul Islam, Afshin Somani
Department of Computer Science and Software Engineering
Concordia University,
Montreal, Quebec, Canada

*Abstract— This paper reports an empirical study on refactoring activity in three Java software systems. We investigated some questions on refactoring activity, to confirm or disagree on conclusions that have been drawn from previous empirical studies. Unlike previous empirical studies, our study found that it is not always true that there are more refactoring activities before major project release date than after. In contrast, we were able to confirm that software developers perform different types of refactoring operations on test code and production code, specific developers are responsible for refactorings in the project, refactoring edits are not very well tested. Further, floss refactoring is more popular among the developers, refactoring activity is frequent in the projects, majority of bad smells once occurred they persist up to the latest version of the system. By confirming assumptions by other researchers we can have greater confidence that those research conclusions are generalizable.*

## I. INTRODUCTION

Refactoring improves the structure of the software in a way that it does not alter the external behavior of the code [1]. It is believed that refactoring improves software quality, developer's productivity, maintainability and understandability of the software systems [2]. Many believe that lack of refactoring causes technical debt, which might result in increased maintenance costs [3]. The empirical studies conducted previously have analyzed important aspects of refactoring practice. In this paper, we investigated seven questions on refactoring activity, to confirm or disagree on conclusions that have been drawn from previous empirical studies.

We inspected the refactoring history of three well known projects namely JFreeChart, JEdit and JMeter and investigated 7 research questions related to refactoring practice. These questions are:

RQ1: Is refactoring interleaved with other types of maintenance activity (floss refactoring) or is it performed in a periodic fashion (root-canal refactoring)?

RQ2: Are there adequate tests for refactoring edits in practice?

RQ3: Do software developers perform different types of refactoring operations on test code and production code?

RQ4: Which developers are responsible for refactorings in the project?

RQ5: Is there more refactoring activity before major project releases than after?

RQ6: Is refactoring activity frequent in the project?

RQ7: Does the number of code smells increase over time and are the problems being solved during the evolution of the software?

The answers to these research questions will help to better understand the actual refactoring practice.

The remainder of this paper is organized as follows. Section II summarizes related work. Section III describes our study approach; Section IV presents our results; Section V discusses the threats to the validity of the study; Section VI concludes with the direction of future work.

## II. RELATED WORK

Nikolaos Tsantalis and Victor Guana [4] conducted a multidimensional empirical study on refactoring activity. The results from this study showed that refactorings applied on production code mainly focuses on design improvements whereas refactorings applied on test code focuses on internal organization of classes corresponding to each project. The study also found that single developer acted as a refactoring manager. Further, they found the amount of refactorings applied on a project before release was predominantly higher when compared with the refactorings applied on the project after release. Additionally, intense refactoring activity was observed during the testing periods of the project and also found interesting facts behind the purpose of applied refactorings.

Murphy-Hill et al. [5] used the version history of Eclipse code base from CVS repository to investigate the refactoring practices followed by the developers. They concluded that refactoring is used frequently to add a new feature or to fix a bug (floss refactoring), comments do not provide with useful information and the percentage of low-level and medium-level refactorings (sub method level) is higher compared with the high-level refactorings (changing the signature of class, methods). Moreover, the experiment found that nearly 90% of the refactorings were applied manually without using any refactoring tool.

Kim et al. [6] used fine grained evolution history of the projects namely Eclipse JDT Core, JEdit and Columba and investigated the role of API level refactorings. They concluded that API level refactorings increase the bug fixes, time taken to fix a bug after applying API level refactorings is shorter than before and refactorings at API level occur more before the major software releases.

Kim et al. [7] investigated refactoring benefits and challenges at Microsoft through survey, interviews with professional software engineers and version history data analysis. The results of the study showed that the refactoring definition in practice differs from academic definition of behavior preserving program transformation. The study also found that refactoring involves lot of costs and risks and various support is necessary to implement refactoring beyond automated refactoring within IDEs.

Rachatasumrit and Kim [8] investigated the impact of refactoring edits on regression testing using the version history of three open source software systems namely: JMeter, XMLSecurity, and ANT. The results of this study showed that only 22% of refactorings are properly tested. The study also suggested that half of failed affected tests include refactoring edits.

Gabriele Bavota and Bernardino De Carluccio [9] found that there is no significant difference in the proportion of classes involved in bug-fix inducing changes. Some refactorings like the Pull up method or Inline Temp were identified which can induce more bug fixes. Considering the source and the target classes, it was identified that the Move Field and the Move method are prone to induce errors in target classes, while the Replace Method with Method Object and Pull Up method induce bugs in the source classes. The results helped to conclude that the fault induced by refactoring is about 15%.

A. Chatzigeorgiou and A. Manakos [10] investigated the evolution of bad smells in object-oriented code. F. Khomh, M. Di Penta and Y.-G. Guéhéneuc [11] have done an exploratory study of the impact of code smells on software change-proneness. S. Olbrich, D. S. Cruzes, V. Basili and N. Zazworka [12] studied two open source software systems to find out the evolution and impact of code smells.

## III. STUDY APPROACH

We selected JFreeChart, JEdit and JMeter as our study subjects because they have high quality change logs. We have investigated 10 versions of JFreeChart, JEdit and 9 versions of JMeter. JFreeChart is a free chart library for the Java platform. It supports bar charts, pie charts, line charts, time series charts, scatter plots, histograms, simple Gantt charts, Pareto charts, bubble plots, dials, thermometers and more [13]. JEdit is a programmer's text editor written in Java. It uses the Swing toolkit for the GUI and can be configured as a rather powerful IDE through the use of its plugin architecture [14]. JMeter is a Java application designed to load test functional behavior and measure performance. It was originally designed for testing Web Applications but has since expanded to other test functions [15].

*Step 1: Identifying Refactorings*

In order to identify the refactoring edits in the selected three software systems we have used Ref-Finder [16] refactoring reconstruction tool, which is available as an Eclipse plugin. Ref-Finder is able to detect 63 different kinds of refactorings. With the help of this tool we first compared the base version with the corresponding changed version and continued this process until all the versions are inspected and tabulated the type of refactorings along with the affected classes in the source code. In a case study it is found that Ref-Finder was able to detect refactoring operations with an average recall of 95% and an average precision of 79% [17].

*Step 2: Identifying the Test Coverage of Refactored Code*

To investigate whether the identified refactorings have test coverage or not we have used Eclemma [18] tool. Eclemma is a free Java Code Coverage tool for Eclipse, which is openly available as an Eclipse plugin. Eclemma highlights the source code with green color if the code has test coverage. We have run Eclemma on our three software systems JFreechart, JEdit and JMeter and have documented the results.

*Step 3: Identifying the authors and specific date of the refactorings*

From the Ref-Finder tool results we have learned in which classes those refactorings have been applied. Than we have checked the commit history of those classes. For JFreeChart and JEdit projects we have used TortoiseSVN subversion client, which is also integrated into Eclipse IDE. Then with the help of commit logs we have found the author of particular refactoring and the specific date when that refactoring was actually applied. But to analyze the commit history of JMeter we have used Git repository and performed similar analysis as done by the other two projects.

*Step 4: Identifying Bug fixes and New Features Implementation*

To identify bug fixes and new features implementation of JEdit and JFreechart projects we have used the website sourceforge.net. This website keeps track of current bugs and also when bugs have been fixed. For our study we have only considered those bugs, which have closed and fixed status. When new features have been implemented this website also keeps track of that. But for JMeter we have checked Bugzilla for issue tracking.

*Step 5: Identifying Code Smells*

To identify the code smells we used Eclipse JDeodorant [19] plugin. JDeodorant employs a variety of unique methods and techniques in order to identify code smells and suggests the appropriate refactorings to resolve them. For our study we emphasized only on God Class, Feature Envy, and Type Checking code smells. Initially, we have inspected code smells for the oldest version. Then for each project we have checked whether the code smells we have found exist in the next consecutive versions.

## IV. RESULTS

### RQ1: Is refactoring interleaved with other types of maintenance activity (floss refactoring) or is it performed in a periodic fashion (root-canal refactoring)?

According to Murphy-Hill and Black programmers use two tactics when they perform refactoring: floss refactoring and root-canal refactoring [20]. If programmers do refactorings to reach a specific goal like adding a new feature or fixing a bug it is called floss refactoring. To maintain healthy code floss refactoring is used. On the other hand, root canal refactoring is used to correct unhealthy code. Root canal refactoring does not change the semantics of the program. For convenience, we decided that a programming session will be the time period of one week. Than for each examined project, we focused only on those weeks when programmers have done a lot of refactorings. In JFreeChart, we have found 26 floss refactorings and 3 root-canal refactorings. For JMeter project, we have discovered 49 floss refactorings but only 7 root-canal refactorings. In JEdit, we have found 17 floss refactorings and 2 root-canal refactorings. From our analysis it is clear that floss refactoring is more popular among the developers. Overall, our study confirms the claim of Murphy-Hill and Black that floss refactoring is a more common practice [5].

### RQ2: Are there adequate tests for refactoring edits in practice?

We investigated whether the source code involved in refactoring has test coverage or not with help of Eclemma tool. In our study we have found that refactoring edits are not very well tested. Out of 389 detected refactorings in JFreeChart, only 117 refactorings have test coverage. So, for JFreeChart 30% of refactorings are tested. In JMeter, out of 670 detected refactorings only 188 refactorings have test coverage. So, for JMeter 28% of refactorings are tested. Surprisingly for JEdit there was no test code in the project. So, JEdit project does not have any test coverage. Overall, our study found same kind of results as previous empirical study done by Rachatasumrit and Kim [8].

### RQ3: Do software developers perform different types of refactoring operations on test code and production code?

We have found that software developers perform different types of refactoring operations on test code and production code. In JFreeChart, test code is placed within package tests.junit. We detected total 389 refactorings in JFreeChart. Among them 304 (78%) were production code refactorings and 85 (22%) were test code refactorings. In JMeter we noticed that test code is placed within a specific folder called test. We detected total 670 refactorings in JMeter. Among them 600 (90%) were production code refactorings and 70 (10%) were test code refactorings. Surprisingly for JEdit there was no test code in the project. So all 255 detected refactorings were production code refactorings.

Table 1 shows the number of detected refactorings on production code and test code for each examined project. In JFreeChart, the most frequent refactoring operation in test code was Introduce Assertion. The goal of this refactoring was to make the assumption explicit with an assertion. For JMeter, we found that the three most dominant refactoring operations in test code were Replace Magic Number With Constant, Add Parameter and Rename Method. On the other hand, for each examined project the most popular refactoring operations in production code were Consolidate Duplicate Conditional Fragments, Introduce Explaining Variable, Extract Method, Add Parameter and Replace Method With Method Object. The goal of the applied refactorings on production code was often to remove the duplicate code, make long methods shorter and to improve the understandability of the code. Finally, refactoring operations Replace Parameter With Method, Push Down Method, Replace Data With Object, Self-Encapsulate Field, Replace Constructor With Factory Method, Preserve Whole Object, Introduce Explaining Method, Push Down Field, Replace Temp With Query and Form Template Method were rarely applied in the projects. Overall, our study result confirms the claim of Tsantalis and Guana that software developers perform different types of refactoring operations on test code and production code [4].

|  | Production Code | | | Test Code | | |
| --- | --- | --- | --- | --- | --- | --- |
| Refactoring Type | JFreeChart | JEdit | JMeter | JFreeChart | JEdit | JMeter |
| Consolidate Duplicate Conditional Fragments | 47 | 40 | 70 | 1 | 0 | 4 |
| Introduce Explaining Variable | 37 | 21 | 42 | 0 | 0 | 6 |
| Extract Method | 28 | 30 | 22 | 0 | 0 | 5 |
| Add Parameter | 20 | 25 | 68 | 1 | 0 | 11 |
| Replace Method With Method Object | 21 | 14 | 43 | 5 | 0 | 3 |
| Remove Parameter | 15 | 13 | 55 | 0 | 0 | 3 |
| Inline Temp | 22 | 14 | 21 | 0 | 0 | 3 |
| Remove Assignment To Parameters | 37 | 8 | 14 | 1 | 0 | 1 |
| Rename Method | 5 | 11 | 21 | 5 | 0 | 8 |
| Replace Magic Number With Constant | 7 | 10 | 90 | 1 | 0 | 13 |
| Introduce Assertion | 3 | 6 | 5 | 62 | 0 | 1 |
| Consolidate Conditional Expression | 16 | 7 | 17 | 1 | 0 | 0 |
| Move Method | 3 | 19 | 23 | 3 | 0 | 3 |
| Replace Nested Conditional Guard Clauses | 19 | 2 | 4 | 0 | 0 | 1 |
| Move Field | 3 | 7 | 24 | 0 | 0 | 2 |
| Extract Interface | 7 | 0 | 21 | 0 | 0 | 1 |
| Remove Control Flag | 3 | 4 | 9 | 1 | 0 | 0 |
| Introduce Null Object | 5 | 0 | 7 | 1 | 0 | 2 |
| Inline Method | 1 | 6 | 16 | 0 | 0 | 1 |
| Replace Exception With Test | 3 | 0 | 4 | 0 | 0 | 0 |
| Extract Superclass | 1 | 2 | 3 | 0 | 0 | 0 |
| Separate Query From modifier | 0 | 3 | 1 | 3 | 0 | 1 |
| Remove setting method | 0 | 3 | 3 | 0 | 0 | 0 |
| Hide Delegate | 1 | 1 | 3 | 0 | 0 | 0 |
| Introduce local extension | 0 | 0 | 5 | 0 | 0 | 0 |
| Replace parameter with method | 0 | 0 | 2 | 0 | 0 | 0 |
| Push down method | 0 | 0 | 2 | 0 | 0 | 0 |
| Replace data with object | 0 | 0 | 2 | 0 | 0 | 0 |
| Self-Encapsulate field | 0 | 0 | 1 | 0 | 0 | 0 |
| Replace constructor with factory method | 0 | 0 | 1 | 0 | 0 | 0 |
| Preserve Whole Object | 0 | 2 | 0 | 0 | 0 | 0 |
| Introduce Explaining method | 0 | 2 | 0 | 0 | 0 | 0 |
| Push Down Field | 0 | 2 | 0 | 0 | 0 | 0 |
| Replace Temp with Query | 0 | 1 | 0 | 0 | 0 | 0 |
| Form Template Method | 0 | 2 | 1 | 0 | 0 | 1 |

**Table 1: Number of detected refactorings on the production and test code**

**RQ4: Which developers are responsible for refactorings in the project?**

The JMeter project has total 23 developers. But refactorings are performed by only 5 developers. Figure 1 shows the percentage of refactoring activities performed by JMeter committers. The top two refactoring contributors are Sebastian Bazley (55%) and Philippe Mouawad (40%) and they are also top two committers with 65% and 15% of the commits respectively. The JEdit project has total 187 developers although most of them are not big contributors. They have at least one commit. But refactorings are performed by only 20 developers. Figure 2 shows the percentage of refactoring activities performed by JEdit committers. The top two refactoring contributors are Matthieu Casanova (42%) and Alan Ezust (37%) and they are also top two committers with 27% and 34% of the commits respectively. For JFreeChart project there are only 3 developers. Figure 3 shows the percentage of refactoring activities performed by JFreeChart committers. In this project top refactoring contributor is David Gilbert (99%). He is also top committer in the project with more than 99% of commits. Overall, our study result confirms the claim of Tsantalis and Guana that most of the refactorings are performed by specific developers that usually have a key role in the management of the project [4].

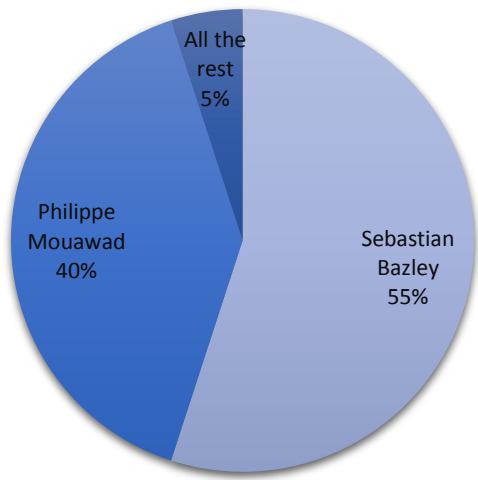

**Figure 1: JMeter Refactoring Contributors**

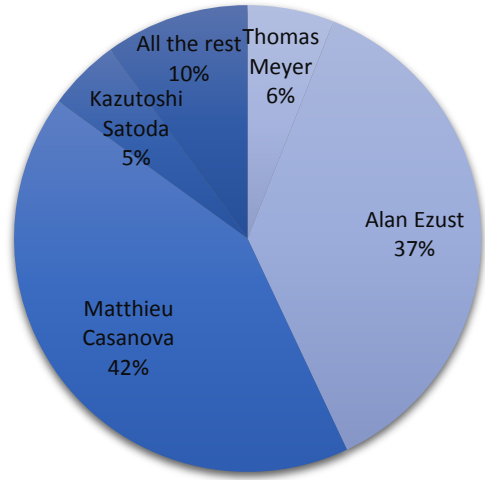

**Figure 2: JEdit Refactoring Contributors**

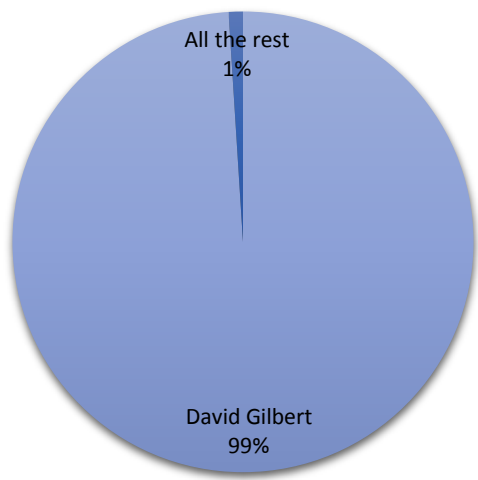

**Figure 3: JFreeChart Refactoring Contributors**

**RQ5: Is there more refactoring activity before major project releases than after?**

To investigate whether refactoring activity is more before major project release or after we have selected windows of 80 days around each release date. Here each window was divided in two groups of 40 days to analysis the refactoring activity before and after a release point. Than we counted for that 80 days period, how many refactorings are performed in each day. Although we have investigated 10 releases of JFreeChart and JEdit but to answer this specific question we discarded those project release versions, which have time overlap. Figures 4, 5 and 6 present the refactoring activity around the release dates of JEdit (8 releases), JFreeChart (8 releases) and JMeter (9 releases) respectively.

In the JEdit project, Figure 4 shows that refactoring activity peaked 2 days before the release day. For JFreeChart project, Figure 5 shows constant refactoring activity in the last 2 weeks of the project before the release. Moreover, the refactoring activity peaked 1 month before the release day. We also observed that there was also refactoring activity after the release. For JMeter we got contradictory results because there was more refactoring activity after the release. Figure 6 shows constant refactoring activity after the release day in multiple release points. Furthermore, the refactoring activity peaked 35 days after the release day. Although, we also observed that there was a lot of refactoring activity in the period between 40 and 5 days before the release day. We also noticed that most refactorings that were performed after release day was because, the developers wanted to fix bugs. Overall, our study not always found that there is more refactoring activity before major project releases than after as claimed by Tsantalis and Guana [4].

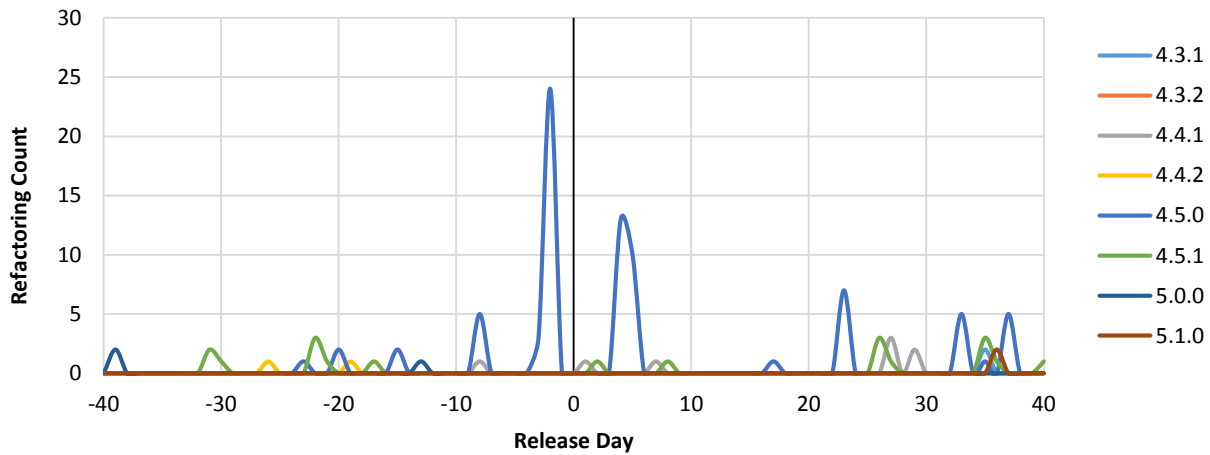

**Figure 4: Refactoring Activity Comparison (Release ) – JEdit (40 days before and after release)**

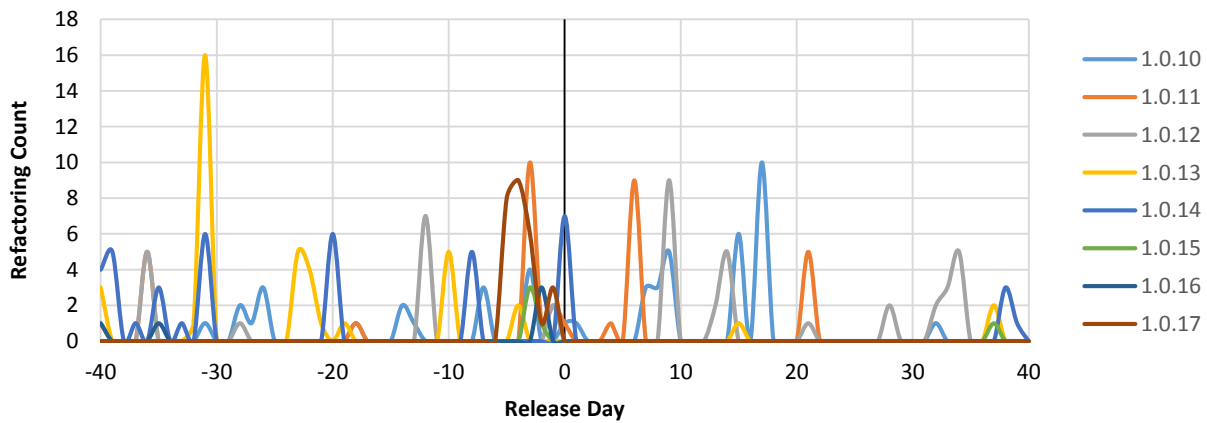

**Figure 5: Refactoring Activity Comparison (Release ) – JFreeChart (40 days before and after release)**

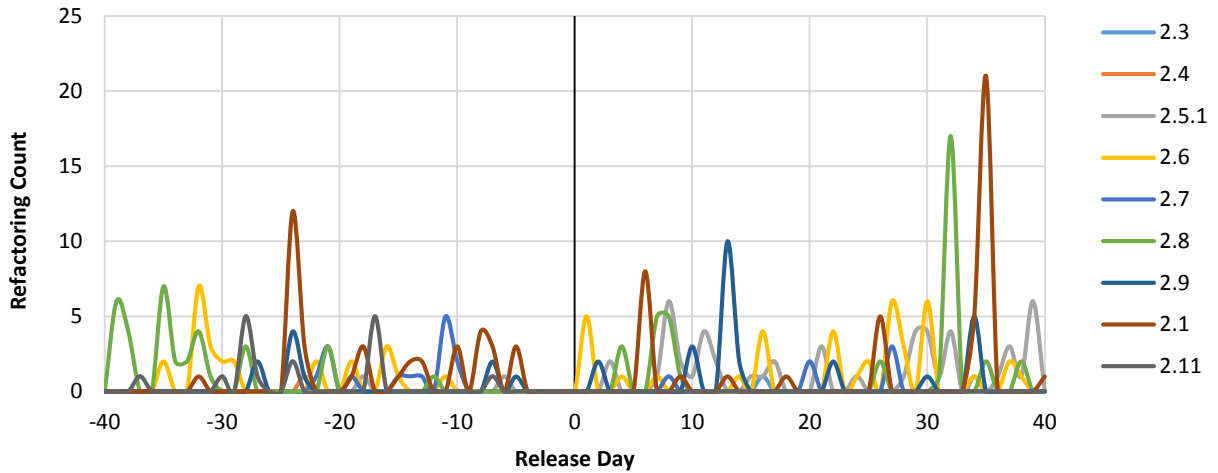

**Figure 6: Refactoring Activity Comparison (Release ) – JMeter (40 days before and after release)**

**RQ6: Is refactoring activity frequent in the project?**

From our study, we have found that refactoring activity is frequent in the project. Figures 7, 8 and 9 present refactoring activity of JFreeChart, JEdit and JMeter respectively. In JFreeChart, Figure 7 shows that refactoring activity was at peak in April 2009. It is also clear from the graph that not many refactorings have been performed in recent years in this project. On the other hand, in JEdit, Figure 8 shows that refactoring activity was at peak in January 2012. In JMeter, Figure 9 shows that refactoring activity is increasing in recent years. Moreover, we noticed that in this project refactoring activity was at peak in August 2013. In each of the examined project we also noticed that when there were a lot of commits in the project there were also a lot of refactorings. Overall, our study confirms the claim of Murphy-Hill and Black that refactorings are frequent in the project [5].

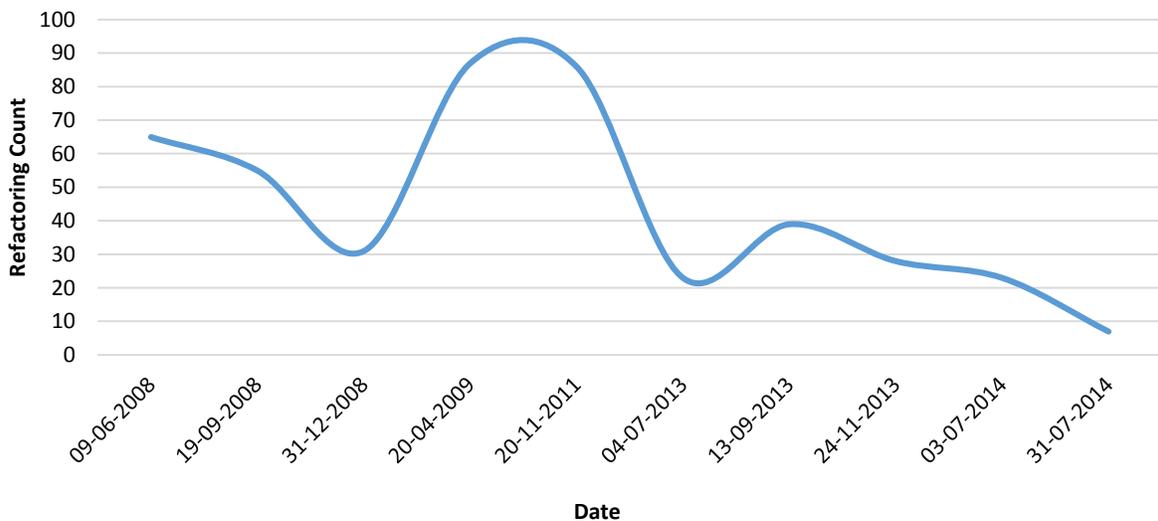

**Figure 7: Refactoring Activity of JFreeChart**

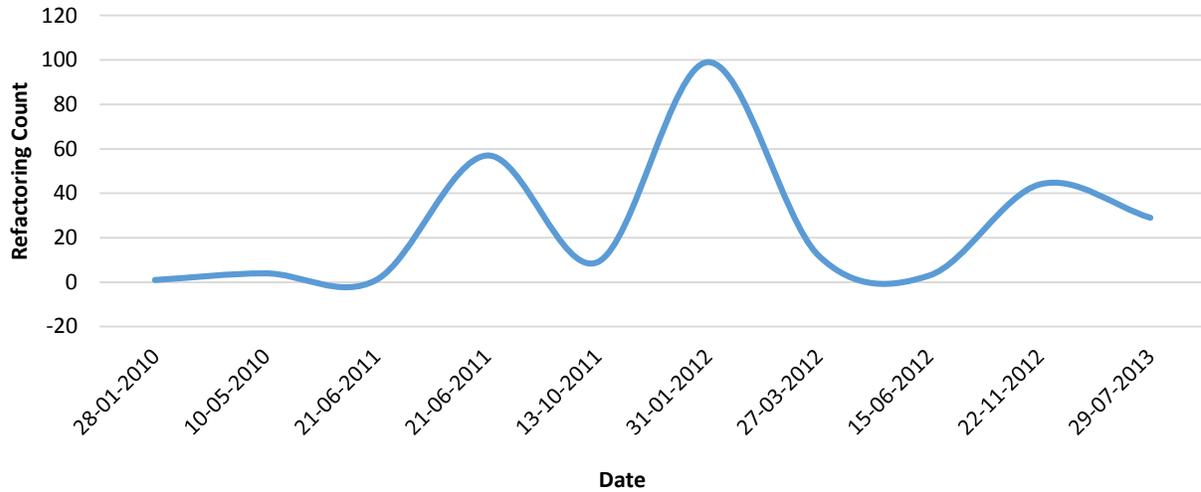

**Figure 8: Refactoring Activity of JEdit**

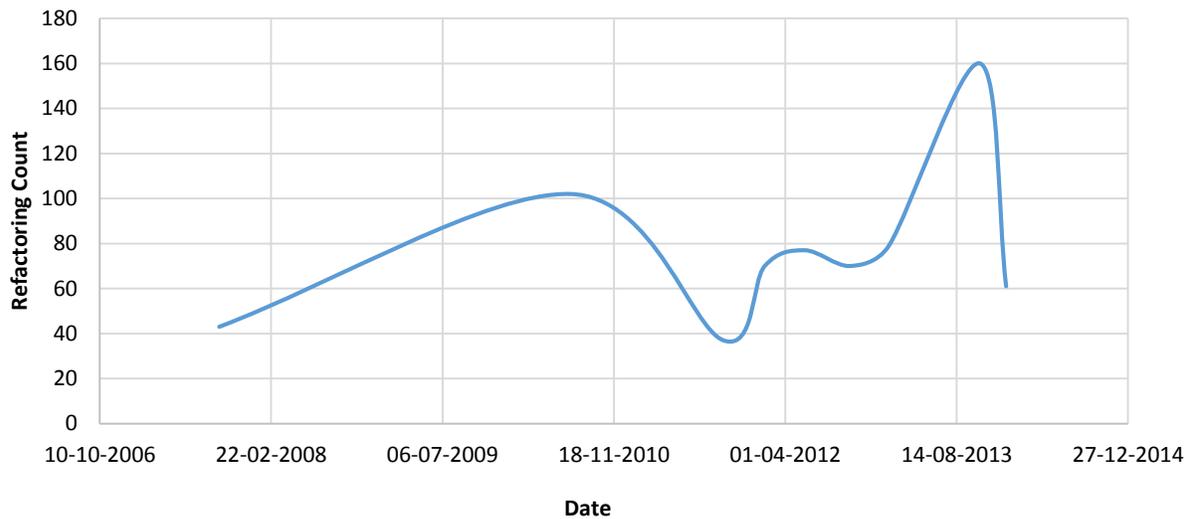

**Figure 9: Refactoring Activity of JMeter**

**RQ7: Does the number of code smells increase over time and are the problems being solved during the evolution of the software?**

We have considered three types of code smells here which are God Class, Feature Envy and Type Checking. In JFreeChart version 1.0.10, we have found 137 God Class, 44 Feature Envy and 5 Type Checking code smells. On the other hand, for JEdit version 4.3.1 we have found 178 God Class, 219 Feature Envy and 86 Type Checking code smells. For JMeter version 2.1 we have found 180 God Class, 286 Feature Envy and 16 Type Checking code smells. After the analysis we found that the majority of bad smells once occurred they persist up to the latest version of the system. We also found that number of code smells increases as the system evolves. Because, functionalities are added in every new release and there is no systematic maintenance of open source software systems the result is expected. Overall, our study found same results as previous empirical study conducted by Chatzigeorgiou and Manakos [10].

## V. THREATS TO VALIDITY

All refactoring operations identified by the tool Ref-Finder were not manually analyzed through source code inspection. Thus, the validity of our study is threatened by the problems related to false positives. Although our refactoring detection tool Ref-Finder has very high accuracy, it may miss some actual refactorings that have been applied in the projects. Here, the validity of our study is threatened by the presence of false negatives. Further, we only considered three open source Java projects namely JFreeChart, JEdit and JMeter and thus the results of the study may not be generalizable to other programming languages. Our study results may be strongly influenced by a few developers' practices.

## VI. CONCLUSIONS

In this empirical study, we investigated seven questions addressing refactoring activity in three open source Java software systems. Our observations includes, floss refactoring is more common, refactoring edits are not very well tested. Nevertheless our research also addressed that refactoring operations on test code and production code are different, and only few developers are responsible for refactoring activities among all the contributors. Additionally we have also found refactoring activities are frequent in the selected projects. Moving further we have also noticed that it is not always true that there are more refactoring activities before major project release date than after. Finally from our manual inspection we have noticed that majority of bad smells once occurred they persist up to the latest version of the system. Future research should investigate the motivation behind the applied refactorings and find the effect of the detected refactorings on the design quality and maintainability of the projects.